\documentclass[sigconf,nonacm=true,authorversion=true]{acmart}

\usepackage[utf8]{inputenc}
\usepackage[T1]{fontenc}
\usepackage{graphicx}
\usepackage{grffile}
\usepackage{longtable}
\usepackage{wrapfig}
\usepackage{rotating}
\usepackage[normalem]{ulem}
\usepackage{amsmath}
\usepackage{textcomp}
\usepackage{capt-of}
\usepackage{hyperref}
\usepackage{listings}

\setcopyright{acmcopyright}
\copyrightyear{2018}
\acmYear{2018}
\acmDOI{10.1145/1122445.1122456}

\acmConference[Woodstock '18]{Woodstock '18: ACM Symposium on Neural
  Gaze Detection}{June 03--05, 2018}{Woodstock, NY}
\acmBooktitle{Woodstock '18: ACM Symposium on Neural Gaze Detection,
  June 03--05, 2018, Woodstock, NY}
\acmPrice{15.00}
\acmISBN{978-1-4503-XXXX-X/18/06}

\begin{document}

\newcommand{\stereo}{\textsc{stereo}}
\newcommand{\flow}{\textsc{flow}}
\newcommand{\convolution}{\textsc{convolution}}
\newcommand{\descriptor}{\textsc{descriptor}}
\newcommand{\X}{$\times$}
\newcommand{\TODO}[1]{#1}
\newcommand{\HLSSTUFF}[1]{#1}

\title{HWTool: Fully Automatic Mapping of an Extensible C++ Image Processing Language to Hardware}
\author{James Hegarty}
\affiliation{%
  \institution{Facebook}
}

\author{Omar Eldash}
\affiliation{\institution{University of Louisiana at Lafayette}}

\author{Amr Suleiman}
\affiliation{\institution{Facebook}}

\author{Armin Alaghi}
\affiliation{\institution{Facebook\vspace{20pt}}}

\begin{abstract}
Implementing image processing algorithms using FPGAs or ASICs can improve energy efficiency by orders of magnitude
 over optimized CPU, DSP, or GPU code. These efficiency improvements
are crucial for enabling new applications on mobile power-constrained devices, such as cell phones or
AR/VR headsets. Unfortunately, custom hardware is commonly implemented using a waterfall process 
with time-intensive manual mapping and optimization phases. 
Thus, it can take
years for a new algorithm
to make it all the way from an algorithm design to shipping silicon. Recent improvements in hardware 
design tools, such as 
C-to-gates High-Level Synthesis (HLS), can reduce design time, but still require manual 
tuning from hardware experts.

In this paper, we present HWTool, a novel system for automatically mapping image 
processing and computer
vision algorithms to hardware. Our system maps between two domains: 
HWImg, an extensible C++ image processing
 library containing common image processing and parallel computing operators, 
and Rigel2, a library
of optimized hardware implementations of HWImg's operators and backend Verilog compiler. 
We show how to automatically compile HWImg to Rigel2,
by solving for interfaces, hardware sizing, and FIFO buffer allocation. 
Finally, we map full-scale image processing applications like 
convolution, optical flow, depth from stereo, and feature descriptors to FPGA using our system. 
On these examples, HWTool requires on average only 11\% more FPGA area than hand-optimized designs (with manual FIFO allocation), 
and 33\% more FPGA area than hand-optimized designs with automatic FIFO allocation\HLSSTUFF{, and performs similarly to HLS}.
\end{abstract}

\maketitle

\section{Introduction}
\label{sec:org070c6bf}

\begin{figure}
  \includegraphics[width=240pt]{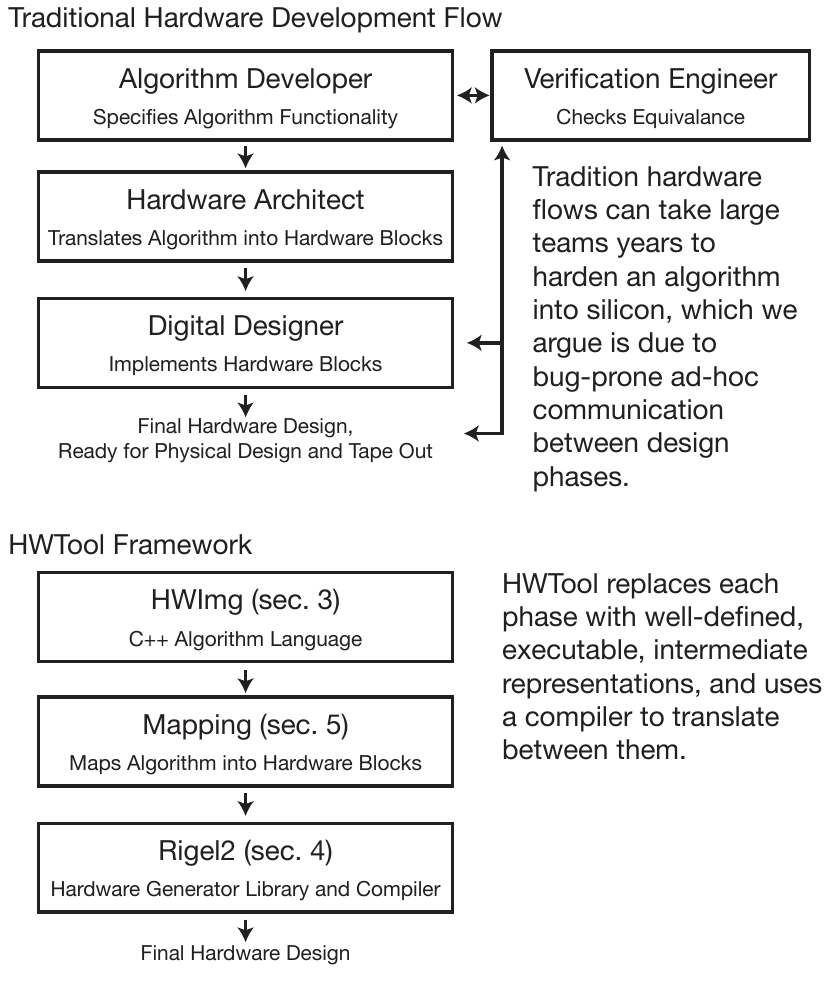}
  \vspace{-20pt}
\label{orgcd731bd}
\end{figure}

Prior work has shown that implementing image processing algorithms on FPGAs or ASICs can yield a 500\X~ 
improvement in energy efficiency over CPUs \cite{inefficiency}. This efficiency improvement is becoming 
increasingly
necessary as mainstream computing moves towards low-power mobile devices, such as cell phones, AR/VR 
headsets, and robots. The latest research from image processing and computer vision communities
simply cannot run as software on these platforms, because it would consume too much power
 or exceed thermal limits. 

Unfortunately,
hardening the latest image processing research into custom hardware
is out of reach for most researchers and many companies: the cost of designing and implementing the
hardware is too high, and the development time is too long. Bringing the latest image processing and
computer vision research to a wide userbase will require us to decrease the cost and development 
time of custom hardware.

Hardware design requires implementing
the desired algorithm multiple times, each to satisify different requirements of the development process.
First, the algorithm is implemented as high-level software, to demonstrate that the approach can solve the desired
vision or image processing problem. Next, a hardware architect translates this high-level code
into precisely-specified hardware blocks and makes optimization decisions such as choosing lower integer precision.
Often, they also
implement a software simulation of the precise hardware architecture for reference and testing.
Finally, the desired hardware architecture is implemented again as Verilog to be synthesized.

Each of these manual implementation and optimization steps is an opportunity
to introduce bugs or miscommunications. Furthermore, the long cycle between algorithm development
and hardware implementation means that optimization opportunities discovered at the hardware level
may never have a chance to propagate back to the algorithm design.

Given the opportunities for improvement, hardware design tools have become an
active area of research in recent years. Researchers have proposed replacements for Verilog
 \cite{chisel, magma}, C-to-gates High-Level Synthesis (HLS) tools, which allow a subset
of C to be compiled into hardware \cite{autoesl, vivado}, and hardware-targeted Domain Specific 
Languages
(DSLs) \cite{halidehls, darkroom}. Unfortunately, while these tools
help with individual stages in the hardware design process, none of them help organize and
optimize the full hardware flow from algorithm development to final silicon. Coordinating between
hardware and software development teams, and supporting downstream verification efforts remain open
problems in all of these tools.

In this paper, we present HWTool, an extensible compiler and framework for fully-automatic mapping
of image processing and computer vision code to custom hardware. The HWTool framework and compiler
consists of multiple Intermediate Representations (IRs) that map to the job functions of hardware teams. HWTool
consumes algorithm code written in HWImg, our C++ image processing library
(sec. \ref{sec:org88da7f4}).
HWTool's
mapper (sec. \ref{sec:orgda3d9a8}) finds the locally minimal cost hardware implementation 
of each HWImg operator at each point in the pipeline. 

For HWTool's hardware backend, we significantly improved Rigel to create Rigel2 \cite{rigel}. 
Rigel is a simple hardware IR and standard library of 
\textit{hardware generators}.
Hardware generators produce optimized Verilog for each operator from a set of fixed configuration options.
For example, Rigel's image cropping generator would produce
a Verilog module to crop a chosen number of pixels from a chosen image size.
Rigel2's IR allows the compiler to analyze throughput and interface requirements for each intermediate in the
pipeline, which aids HWTool's mapper
in choosing the most optimal hardware implementation for each operator. Finally, we show how to
solve for FIFO buffer allocation in Rigel2, enabling a fully-automated flow from C++ to hardware
(sec. \ref{sec:org3e5493f}).

A key goal for this project was creating a tool that can be practically used by hardware design 
teams.  HWTool is not intended to replace hardware designers - instead, it is a framework designed to make these teams work more effectively. 
We believe our system must meet the following goals to be successful:
\begin{itemize}
\item \textbf{Efficiency:} Implementing algorithms in fixed-function hardware only makes sense if \textit{extreme} efficiency is 
  required.
  HW design tools that do not approach
the efficiency of hand-tuned 
hardware are not useful, because a large performance regression would make custom hardware
not worth the effort.
\item \textbf{Flexibility and Extendability:} Domain Specific Languages (DSLs) can expose convenient programming models
and perform impressive optimizations, however they are often limited in functionality and hard to extend.
Instead, we must provide a flexible, extendable 
framework that can support almost the full range of hardware expressible in a general-purpose language 
like Verilog and allow for functionality to be added later.
\item \textbf{Interoperability:} Common hardware blocks like caches, camera interfaces, etc. are 
difficult to implement and verify, and most teams will use existing designs for these components. 
Our system must be able to leverage these existing designs, and work as part of a larger ecosystem of existing
hardware tools.
\item \textbf{Controllability:} Low-level generated hardware designs sometimes must be examined by hand
(e.g., to search for bugs or address physical design problems). 
It's important that the mapping the 
compiler performs is easy for a human to understand, debug, and control. This motivates us to create a 
simple and well-defined model for operations the compiler will perform.
\end{itemize}

This paper makes the following contributions:
\begin{itemize}
\item We present HWImg, a high-level C++ image processing language suitable for use by non-hardware-expert 
algorithm developers, and designed with restrictions to make the hardware mapping problem tractable.
\item We present Rigel2, a hardware description IR which enables mapping of HWImg into hardware
by solving for throughput and interface constraints.
\item We demonstrate how to automatically map HWImg programs into Rigel2 by locally mapping each 
operator to the best matching hardware module, and inserting any necessary conversion at the interfaces.
\item We show how to map Rigel2 to Verilog with no annotation required, by solving for 
FIFO buffering with a scheduling module that allows for bursty modules.
\item We show how these contributions allows HWTool to automatically map four large scale 
image processing 
pipelines to FPGA: convolution, depth from stereo, Lucas-Kanade optical flow, and a simple feature 
descriptor. The resulting FPGA designs use only 11\% more FPGA area than hand-optimized designs (with manual FIFO 
allocation), 33\% more area with fully-automatic FIFO allocation, and similar area to HLS.
\end{itemize}

\section{Background: the Challenges of Mapping to Hardware}
\label{sec:org3768de5}

Compiling and optimizing high-level image processing languages for CPUs and GPUs is a well-studied problem
(for example, the Halide auto-scheduler \cite{halideauto}), whereas it has been less
studied for custom hardware.
This section describe the main compiler problems HWTool must solve to produce good quality hardware. 

\subsection{Sizing Hardware to Meet Throughput}
\label{sec:org477f9fe}

\begin{center}
\includegraphics[width=240pt]{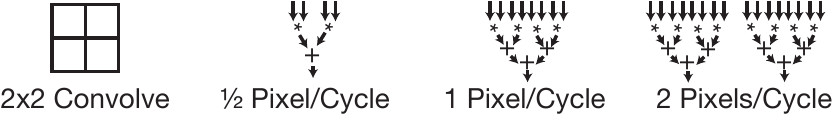}
\end{center}

Hardware compilers must find the minimum set of hardware resources they can allocate to accomplish the given
application at a target performance.
Locally, this decision
is straightforward: the compiler can examine the amount of compute needed in the pipeline, and divide it
by the number of cycles it should take (figure above shows compute needed for convolution at various throughputs).
However, in full-sized pipelines, this decision is more complex. The compiler must understand the throughput impact of all
modules in the pipeline, which may involve tricky data-dependencies or bursty behavior.
Rigel2's IR is explicitly designed to provide this analysis on full pipelines.

\subsection{Latency Matching and Deadlocks}
\label{sec:org34ef8a2}

\begin{center}
\includegraphics[width=240pt]{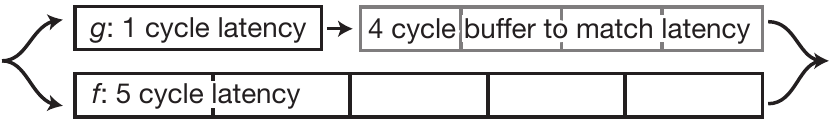}
\end{center}

In pipelines that contain fan-out and reconvergence (above), intermediate values need to be
temporarily stored so that they can be accessed later in the pipeline. Custom hardware typically does not have a
large memory hierarchy that it can spill into, and failing
to allocate sufficient buffering will mean the design will deadlock or perform poorly.

Rigel2 solves this problem by precisely tracking the latency of each
hardware block, and using a solver to optimally solve for the amount of buffering required, as explained
later in section \ref{sec:org4eafc47}.

\subsection{Bursts and Data-Dependent Latency}
\label{sec:orge1658cf}

\begin{center}
\includegraphics[width=240pt]{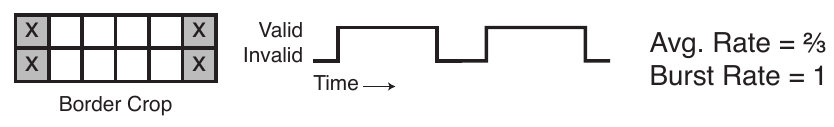}
\end{center}

Many designs include modules with bursty behavior (e.g., crops, above), or data-dependent latency
(e.g., dividers, or cache accesses). 
The compiler analyses HWTool performs, and hardware it generates, must be tolerant of bursts or data dependencies, 
or the scope of hardware it supports will be limited. As with latency matching,
unexpected variability can cause deadlocks or poor performance.

HWTool handles
these challenges by providing a model for characterizing the bursty behavior of individual operators, and 
using this parameter to allocate FIFO buffers to isolate the bursts to that single module in the hardware.

\subsection{Architecture Constraints}
\label{sec:orgb0ce640}

\begin{center}
\includegraphics[width=240pt]{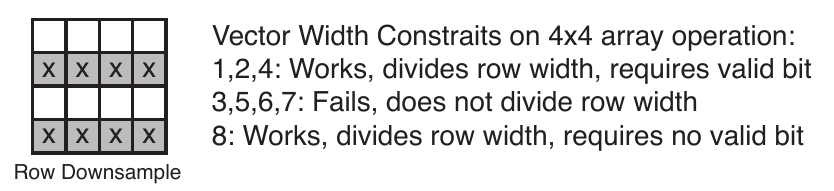}
\end{center}

When mapping high-level operators to hardware, the resulting hardware architecture often has
constraints that must be obeyed. Examples include:
\begin{itemize}
\item Vector width must divide the array size (figure above).
\item Hardware support for some configurations may not exist or may be impossible 
(i.e., some RAM sizes may not be available).
\item The hardware may support a different low-level signaling interface than requested (above).
\end{itemize}

If constraints like these are not dealt with gracefully, most pipelines will fail to map. 
We could attempt to specify these constraints as part of the
\textit{type} of the hardware modules, however we do not know of a practical type system that can support
solving complex constraints like these.

HWTool instead simplifies this problem by supporting mapping to any hardware block that meets or
\textit{exceeds} the constraits, e.g., rounding up to the next largest vector size or using a more complex signaling protocol.
As long as higher-performing implementations are available, this means HWTool can always produce a valid design,
even with difficult or conflicting constraints.
We formally specify the range of options that are allowed and how this can be made reliable in section \ref{sec:org86412f0}.

\section{HWImg: An Extensible C++ Image Processing Library}
\label{sec:org88da7f4}

\begin{figure}
\begin{lstlisting}[language=C++,basicstyle=\small]
class ConvInner:public UserFunction{
public:
ConvInner():UserFunction("ConvInner",
         Array2d(Array2d(Uint(8),2),8,8)){}
Val define(Val inp){
  Val expanded = Map<Map<AddMSBs<24>>>(inp);
  Val products = Map<Mul>(expanded);
  Val sums = Reduce<AddAsync>(products);
  return RemoveMSBs<24>(Rshift<11>(sums));
}};
class ConvTop:public UserFunction{
public:
ConvTop():UserFunction("ConvTop",
              Array2d(Uint(8),1920,1080)){}
Val define(Val inp){
  Val pad = FanOut<2>(Pad<8,8,4,4>(inp));
  Val stencils = Stencil<-7,0,-7,0>(pad[0]);
  Val coeff = RegCoeffs(pad[1]);
  Val convIn = FanIn(Concat(stencils,coeff));
  Val zipped = Map<Zip>(Zip(convIn));
  Val res = Map<ConvInner>(zipped);
  return Crop<12,4,8,0>(res);
}};
\end{lstlisting}
\caption{HWImg is a simple image processing langauge
embedded in C++. Here we show the C++ HWImg code to perform an 8x8 convolution. 
\textit{RegCoeffs} is an external function to load filter coefficients
over the AXI bus.}
\label{fig:HWImgExample}
\end{figure}

\begin{figure}
\raggedright
\textbf{HWImg Types}

\hspace{1em}$T$ := Uint(bits,exp) | Int(bits,exp) | Bits(n) | 

\hspace{3.5em}Float(exp,sig) | bool |

\hspace{3.5em}$T[w]$ | $T[ w, h ]$ | $( T, T, ... )$ \textit{(Arrays and Tuples)}

\hspace{3.5em}$T[ \le w,h]$  \textit{(Sparse Arrays)}

\textbf{HWImg Operators \& Values}

\hspace{1em}$V$ := Input( $T$ ) | \textit{(Function input parameter)}

\hspace{3.5em}Const( $T$, value ) | \textit{(Constants)}

\hspace{3.5em}$f(V)$ | \textit{(Function application)}

\hspace{3.5em}Concat( $V_1, V_2, ...$ )

\textbf{Example HWImg Library Functions}

\hspace{1em}Stencil<$l,r,b,t$> : $T[ w, h ] \rightarrow T[ l+r+1, b+t+1 ][w,h]$

\hspace{2em}\textit{Convert image into image of patches}

\hspace{1em}Crop<$l,r,b,t$> : $T[ w, h ] \rightarrow T[w-l-r,h-b-t]$

\hspace{2em}\textit{Remove outer border of image}

\hspace{1em}Map<$f : T_1 \rightarrow T_2$> : $T_1[w,h] \rightarrow T_2[w,h]$

\hspace{2em}\textit{Convert pointwise function to operate on arrays}
\caption{HWImg's core types and operators. HWImg includes arbitrary-precision ints, floats,
and nested arrays, tuples, and sparse arrays with a maximum size. We also list a few commonly
used template functions in HWImg's standard library. HWImg is monomorphic: all parameters 
must be filled in with
constant values by the template arguments before the functions can be applied.}
\label{fig:HWImgLanguage}
\end{figure}

HWImg is HWTool's C++ image processing language front end. HWImg was designed to simultaneously be
approachable for algorithm developers, but also serve as an input which can be reliably compiled to
hardware. With this in mind, the HWImg library was designed within the following constraints:
\begin{itemize}
\item Arrays or images can only be operated on by fully-parallel array operators. There is no support for loops,
which significantly simplifies dependency analysis and downstream compilation tasks.
\item HWImg is designed to be extendable, and there are few built-in operators. Almost all operations are
performed by generic function calls. New functions can be easily added by developers.
\item HWImg functions are monomorphic: all types and array sizes must be
set at compile time and constant. However, similar to C++ template meta-programming, our front-end syntax sugar has the 
ability to fill in these parameters automatically (see fig. \ref{fig:HWImgExample}). This is necessary, because
types and sizes will be baked into the fixed-function hardware, so these must be constant.
\end{itemize}

A description of HWImg's types, operators, and some common functions in HWImg's standard library in given in
figure \ref{fig:HWImgLanguage}. Example C++ code to implement a convolution using HWImg is given in figure 
\ref{fig:HWImgExample}.

\section{Rigel2 Hardware Backend}
\label{sec:org2d12b12}

\begin{figure}
\raggedright

\textbf{Rigel2 Types}

\hspace{1em}Data Types ($T$) are inherited from HWImg (fig. \ref{fig:HWImgLanguage})

\hspace{1em}Schedule Types (sec. \TODO{4.1}):

\hspace{2em}$S$ := $T$ | $T[ v_w,v_h; w,h \}$ | $S\{ w, h\}$ |

\hspace{4.3em}$T[ v_w,v_h; \le w,h\}$ | $S\{\le w,h\}$

\hspace{1em}Interface Types:

\hspace{2em}$I_s$ := Stream($S$) | ( $I_s$, $I_s$, ... ) | $I_s$[$w$,$h$]

\hspace{2em}$I$ := Static($S$) | $I_s$

\textbf{Rigel2 Operators \& Values}

\hspace{1em}\textit{Input}, \textit{Const}, \textit{Concat}, and \textit{Apply}

\hspace{1em}inherited from HWImg (fig. \ref{fig:HWImgLanguage})

\textbf{Rigel2 Function Properties}

\hspace{1em}Input \& Output Interface type ($I$)

\hspace{1em}Rate, Burstiness, and Latency (sec. \TODO{4.2})

\hspace{1em}Verilog definition string

\caption{Core operators and types in Rigel2. Rigel2 extends HWImg with \textit{Schedule Types}
 to specify vectorized computation, and \textit{Interface Types} to 
describe low-level hardware signaling interfaces. Each Rigel2 function also include scheduling annotations
and a Verilog implementation, which are either derived by the compiler or provided explicitly for imported
Verilog modules.}
\label{fig:Rigel2Language}
\end{figure}

Next, HWTool must map the HWImg code to a hardware description in Verilog.
To accomplish this, we will convert the HWImg pipeline to Rigel2, a novel hardware description language
which can be compiled to Verilog.
Mapping from HWImg to Rigel2 will be discussed later in Section \ref{sec:orgda3d9a8}.

Rigel2 contains key enhancements to the previously-shown Rigel language, which
enable automatic mapping from high-level languages \cite{rigel}. Two features of Rigel2 will be essential
to meeting the goals we set in section \ref{sec:org070c6bf}.
First, Rigel2 can reliably introspect the type and \textit{runtime} throughput of every signal at compile time, which
allows us to automatically specialize each hardware instance to perform optimally in the site where it is needed.
Second, unlike HLS, every module in Rigel2 maps directly to a Verilog module definition. This means that
we can easily import existing Verilog modules (handwritten or generated by another tool) into Rigel2 pipelines,
enabling interoperability with existing code.

A brief overview of Rigel2's types and operators is given in figure \ref{fig:Rigel2Language}. Rigel2 shares its 
core
 data types and operators with HWImg, which makes translating large parts of the language trivial. Rigel2
extends HWImg with some addition hardware-specific annotations. \textit{Interface types} specify the low-level
signaling interface of the hardware. \textit{Static} interfaces are the simplest, and are used for modules that
produce an output in exactly N cycles every cycle. \textit{Stream} interfaces (also known as Handshake or 
Ready-Valid) are more complex and allow for decimation, back-pressure, bursts, etc. 
\textit{Schedule types} are used to enable throughput analysis, and will be discussed later in section \ref{sec:orgfcf0675}.
Rigel2 Functions also include runtime schedule
annotations for rate, burst, and latency, which will be described in section \ref{sec:org4eafc47} and \ref{sec:org3e5493f}.

\subsection{Throughput Tracking}
\label{sec:orgfcf0675}

\begin{center}
\includegraphics[width=240pt]{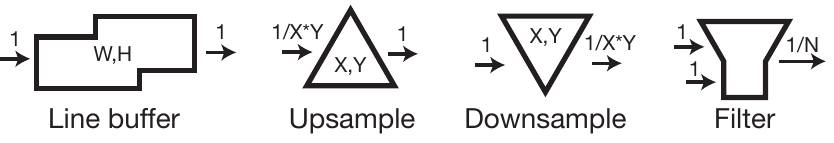}
\end{center}

Rigel supports tracking of hardware rates using the Synchronous Data-Flow (SDF) model \cite{sdf}. In SDF,
hardware is modeled as a graph of modules communicating over data channels.
SDF restricts the behavior of each module. Specifically, the number of output tokens produced by each module must be a fixed ratio of the number of 
tokens consumed, and these annotations must be provided to the scheduler (above). 
For example, a 2-D downsample module would produce \(\frac{1}{4}\) the number of outputs as number of inputs.
SDF rate annotations compose by multiplication (i.e., two downsamples in a row would produce \(\frac{1}{16}\) outputs per input).
By tracking SDF rates throughout the pipeline, the scheduler can statically
 determine the utilization (\% of cycles active) of every interface in the hardware.

Unfortunately, tracking interface utilization alone is insufficient for hardware sizing. Instead, the compiler must track
\textit{throughput}, the number of array elements processed per cycle.
Throughput is a function
of both the utilization of the interface, and the number of elements processed per transaction (which we will
call the vector width, or \(V\)). 
Rigel does not have a way of specifying the vector width of transactions, so throughput cannot be analyzed.

We extended Rigel with \textit{Schedule Types} 
to enable unambiguous vector width, and thus throughput, tracking (fig. \ref{fig:Rigel2Language}). In our syntax, the type 
\(T[ v_w,v_h; w,h \}\) indicates a 2D array operation of size \((w,h)\)  processed at a vector width of 
\((v_w,v_h)\). Vectorized schedule types cannot be nested,
however the special case type \(S\{w,h\}\) is used to build nested non-vectorized operations. Using nesting, the type 
\(T[4,4][2;8,8\}\{256,256\}\) would indicate doing 2 4x4 operations in parallel, and processing the outer 8x8 and 256x256
arrays sequentially. 

\subsection{Buffer Scheduling Model}
\label{sec:org4eafc47}

\begin{figure}[htbp]
\centering
\includegraphics[width=.9\linewidth]{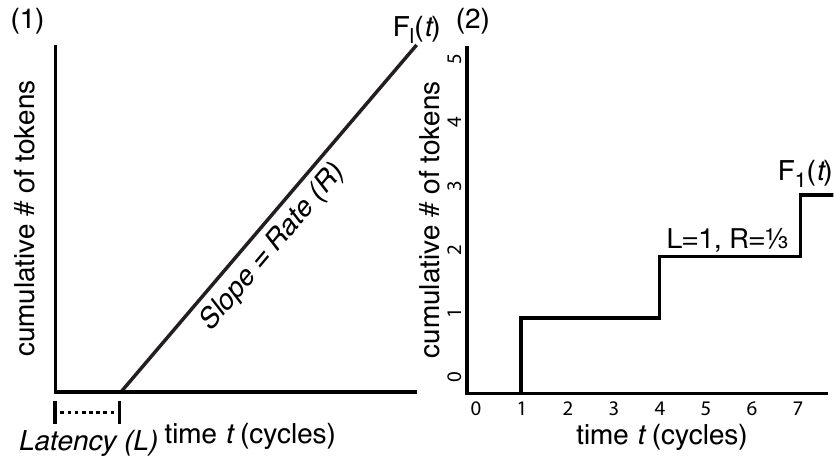}
\caption{\label{fig:orgb6cc931}
Rigel2 include a simple scheduling model based on module latency and rate. (1) plots the number of cumulative tokens that has been produced by a module with a given latency and rate over time. (2) shows a detailed plot of the model over a few cycles, showing how our model discretizes the token count, even in the presence of fractional rates.}
\end{figure}

Rigel supports automatic buffer allocation for single-rate pipelines, but does not support the more common case
of multi-rate pipelines. As a result, many Rigel programs require manual buffer allocation. Here we extend Rigel with a simple scheduling model
for multi-rate pipelines, by extending the core concepts from SDF.

To formally analyze schedules, we will define each module's \textit{token indicator function} \(f(t)\) to be a function 
from cycle \(t\) to \(1\) (for cycles in which 
a token is produced),
or \(0\) (idle). 
Then, we define the function's \textit{schedule trace} \(F(t) = \sum f(t)\), which indicates the cumulative number of
tokens that have been produced or consumed in all cycles up to time \(t\), 

Within our scheduling model, each schedule trace is restricted to the form \(F_L(t) = \max( \lceil (t-L+1)*R \rceil, 0)\).
Rate (\(0 < R \le 1\)) is the number of tokens produced
per cycle, or the SDF rate, and  latency (\(L \ge 0\)) is the number of cycles between when a token is consumed 
and produced by the module.
Figure \ref{fig:orgb6cc931}.1 shows our model's parameterized schedule trace as a function of \(R\) and \(L\). 
The ceiling function discretize the token count, which would otherwise be fractional.
We plot the first few tokens our output at the cycle
level in figure \ref{fig:orgb6cc931}.2. One convenient feature of our model is that the first token is always produced
exactly \(L\) cycles from the start of time.

Schedule traces can be easily shifted in time, which will enable easier analysis of starting and ending
latency.
For convenience, we will define the trace of a function with inputs starting to arrive in
cycle \(s \ge 0\) as \(F_s(t) = F(t-s)\), 
and output trace with outputs arriving in cycle \(s+L\) as
\(F_{s+L}(t) = F(t-s-L)\).

Now, we will use our scheduling model to optimize FIFO buffering in a pipeline.
To ensure correct scheduling, the schedule
trace of each producer must exactly match the trace of its consumers. As explained previously (sec.
\ref{sec:orgfcf0675}), rates \(R\) between all producers and consumers are guaranteed to match
by Rigel's SDF solve, so we do not have to consider this parameter, and only need
to match latencies.

We remark that introducing a FIFO delay buffer of
depth \(d\) in front of a module with start delay \(s\) will delay its output trace from
\(F_{s+L}(t)\) to \(F_{s+L+d}(t)\), with the total size of the FIFO to hold tokens of bitwidth \(b\) equal to \(d*b\).
Thus, formally, given each pair of producers traces \(P_p(t)\) and consumers traces \(C_c(t)\) with respective input delays
\(p\) and \(c\), it must be the case that \(c = p+L_p+d_p\), subject to the constraint \(d_p \ge 0\) (buffers
can not have negative size). Substituting, we get the requirement \(c-p-L_p \ge 0\), with the objective function
\(\sum_{p,c} (c-p-L_p)(b_p)\), which minimizes the amount of buffering required.
This exactly matches the formulation of the register minimization algorithm, which is commonly used to
optimize register allocation in hardware \cite{retiming, darkroom}. 
We found it convenient and sufficiently fast
to solve register minimization using Z3 \cite{z3}, however
this problem also 
has a polynomial solution by reducing to min-cost flow.

\subsection{FIFO Burst Buffering}
\label{sec:org3e5493f}

\begin{figure}[htbp]
\centering
\includegraphics[width=.9\linewidth]{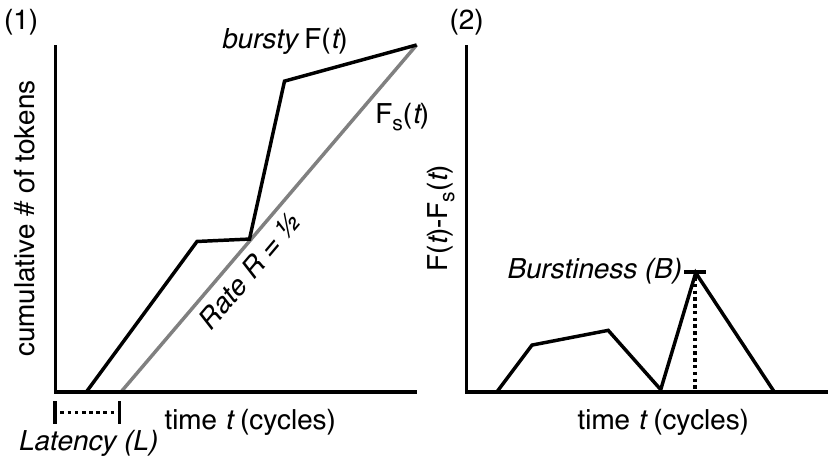}
\caption{\label{fig:orgc0bd8b4}
Bursty modules do not fit directly into the scheduling model presented in section \TODO{4.2}. (1) shows how a schedule trace \(F(t)\) from a bursty module may momentarily exceed the number of tokens in a trace in our scheduling model, \(F_s(t)\). (2) shows how the difference between \(F(t)\) and \(F_s(t)\) can be used to size a FIFO buffer to absorb bursts and fit into our standard model.}
\end{figure}

As motivated in section \ref{sec:orge1658cf}, some important hardware modules have \textit{bursty} behavior: i.e.,
their rate may momentarily exceed their average rate. Bursty behavior can lead to poor
performance if downstream modules only support the average rate, and deadlocks can occur
if the burst fills buffers to the point where draining becomes impossible. 

One solution to prevent
poor performance and deadlocks from bursts is to allocate First-In First-Out (FIFO) buffers around the bursty module, which absorb the bursts
and isolate the rest of the pipeline from them.
We now show how to extend our scheduling model and solver (sec. \ref{sec:org4eafc47})
to support bursty modules. We show in figure \ref{fig:orgc0bd8b4}.1 how a schedule trace \(F_s(t)\) within our scheduling
model compares to \(F(t)\), the module's actual runtime behavior. Even though \(F(t)\) and \(F_s(t)\) converge to the
same average rate at the end of time (\(R=\frac{1}{2}\)), \(F(t)\) has moments where it bursts to \(R=1\), sits idle
 (\(R=0\)), and rates in between. 

We remark that \(F(t)\) has produced more tokens than \(F_s(t)\) in every
cycle, and this can always be guaranteed by setting \(L\) sufficiently large.
Thus, \(F(t)\)'s excess tokens can be temporarily held in a FIFO, and only written out
at the time expected by \(F_s(t)\). We can determine the maximum
size needed for the FIFO by finding the maximum of the excess of \(F(t)\)
relative to \(F_s(t)\). We will call the maximum value attained by \(F(t)-F_s(t)\) the \textit{Burstiness} (\(B\)),
which we plot in figure \ref{fig:orgc0bd8b4}.2.

Each bursty module in Rigel2 must specify its \(L\), \(R\), and \(B\). 
These parameters can often be derived analytically from the expected behavior of the module. 
However, we have often found it most convenient to write
a simulator of the burst behavior (as a function of the cycle), and record \(L\) and \(B\) by fitting a line to the resulting schedule trace. 

Rigel2 also supports data-dependent bursty behavior through a general-purpose 
filter function, which takes in an array and a boolean mask. In these cases, the user needs to explicitly annotate
the expected \(L\) and \(B\) for each filter operator, based on the worst case bursts that they expect to see in 
real-world usage of the pipeline.

\section{Mapping From HWImg To Rigel2}
\label{sec:orgda3d9a8}

\begin{figure}[htbp]
\centering
\includegraphics[width=.9\linewidth]{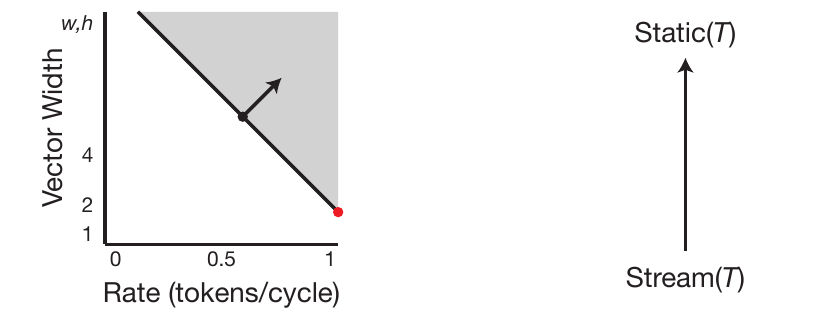}
\caption{\label{fig:orgf378fa4}
Rigel2 supports mapping HWImg operators to either Static or Stream interfaces, and throughputs to a tradeoff space between vector width and rate. The most optimal throughput point is the lowest vector width with a rate of 1 token/cycle (shown in red). This has the smallest vector width, and therefore lowest hardware cost. Static interfaces are preferred over Stream because they are simpler and allow for deeper analysis.}
\end{figure}

Finally, this section describes how to map a high-level program in HWImg to a hardware pipeline
in Rigel2. As explained in section \ref{sec:org3768de5}, mapping is not trivial: it must
correctly size the hardware to meet throughput requirements
 (sec. \ref{sec:org477f9fe}), and also accommodate hardware constraints
 (sec. \ref{sec:orgb0ce640}). Reliable composition of operators in HWImg
requires us to solve these problems consistently, or else various combinations of configurations, throughputs, and
operators will fail to map.

While optimizing for constraints and throughputs globally would lead to the lowest overhead, we think this
would be difficult to solve and hard for the user to comprehend.
Instead, HWTool's has taken an approach where each HWImg operator gets mapped \textit{locally}
to a hardware block that meets \textit{or exceeds} the requirements at that point in the pipeline. Then, we
only have to solve the simpler problem of allowing modules with different (but compatible) interfaces to be composed.
Figure \ref{fig:orgf378fa4} 
specifies the set of allowed type, rate, and vector width substitutions. The key idea
is that a higher throughput or simpler interface can always
be converted to support a lower throughput or more complex interface. 

The first step in mapping is to walk the entire pipeline and determine if a Static or Stream interface is required
(sec. \ref{sec:org75a5ecc}). Following this, the compiler walks the HWImg pipeline a second time, and runs a \textit{mapping function} for
each operator, which returns a Rigel2 module instance that meets or exceeds throughput and rate requirements at that site
(sec. \ref{sec:orgd5a53ee}). 
Finally, interfaces between the mapped Rigel2 modules are converted to match (sec. \ref{sec:org86412f0}).

\subsection{Top-Level Interface Solve}
\label{sec:org75a5ecc}

HWImg functions sometimes get mapped to either to Static or Stream interfaces depending on configuration options and
schedule, so the top-level interface type must be solved for each choice of schedule.
Any pipeline
can be promoted to a Stream interface, however it is desirable to keep a pipeline Static if possible, as this
simpler interface enables more optimal buffer allocation and simplifies some hardware.

In this pre-mapping pass, we send a Static variation of the input type into the input
of the pipeline, and perform mapping and propagation through the pipeline. If at any point
a mapping function returns a Rigel2 function with a Stream interface, we halt and mark the pipeline as Stream.
If all functions get mapped to Rigel2 functions with Static interfaces, we know the top-level input can be Static.

\subsection{Mapping Functions}
\label{sec:orgd5a53ee}

\begin{figure}
\begin{verbatim}
// Reduce is a higher-order operator that applies a
// binary function to tree-reduce an array:
// Reduce( fn : (T,T)->T ) : T[w,h]->T
function ReduceMapper( param, type, rate ):
  binopType = Static( type.over, type.over )
  fn = param.fn:specialize( binopType )
  if fn.latency>0: 
    // fn takes multiple cycles: must be parallel
    return Rigel.Reduce( fn, type.size )
  else:
    tyopt = type:optimize( rate )
    if tyopt.V < tyopt.size: // input is vectorized
      return Rigel.ReduVec(fn, tyopt.size, tyopt.V)
    else: // input is fully parallel
      return Rigel.Reduce( fn, type.size )
\end{verbatim}
\caption{Pseudo-code for the Reduce operator's mapping function. Each operator in HWTool has unique 
requirements to satisfy. For example, Reduce can only perform 
a multi-cycle reduction if the reduction function has zero latency. 
Higher-order operators must recursively map 
the function they operate over (using the \textit{specialize} API).}
\label{fig:reducemapper}
\end{figure}

\textit{Mapping functions} take a HWImg operator and convert it to a Rigel2 generator instance that
that meets or exceeds the throughput and
interface requirements for its location in the hardware pipeline (fig. \ref{fig:orgf378fa4}). 
Mapping functions for operators are provided (1) a set of operator defined input arguments (e.g., downsample 
factor for the downsample operator), and (2) the solved type and rate
 at this point in the pipeline from Rigel2. Example pseudo-code for the mapping function for the Reduce function is given in 
figure \ref{fig:reducemapper}.

Mapping functions must be manually specified for each operator. 
From our experience, mapping cannot be easily
automated: each operator and hardware variant has a unique set of constraints that the mapping function
must satisfy. 

In our implementation, mapping functions are specified in Lua \cite{lua}. HWTool provides
APIs to make writing mapping functions easier, including an API for introspecting and constructing
Rigel2 interface types and rates. A few noteworthy functions seen in figure \ref{fig:reducemapper} are 
\textit{type:optimize},
which returns the interface type that has the lowest valid vector width, and therefore lowest cost
 (the red point in figure \ref{fig:orgf378fa4}), and \textit{HWToolFunction:specialize}, which performs recursive
mapping on another HWTool function to enable implementation of higher-order functions.

\subsection{Automatic Interface Conversion}
\label{sec:org86412f0}

\begin{figure}[htbp]
\centering
\includegraphics[width=240pt]{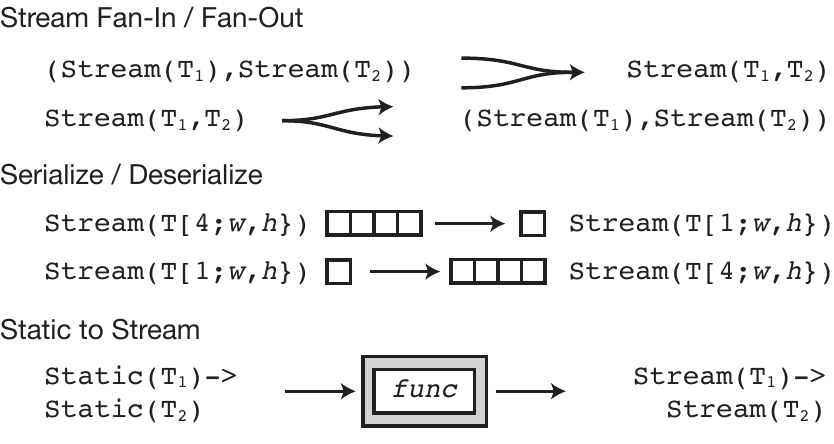}
\caption{\label{fig:org5fb76c4}
HWTool inserts automatic conversions to match interfaces between modules. Fan-In converts tuples of Streams to Streams of tuples (and Fan-Out the opposite). Serialization and De-serialization perform vector width and rate tradeoffs. Finally, Static interfaces can be converted to Stream.}
\end{figure}

Following mapping, each HWImg operator has been converted into a Rigel2 module instance that meets or exceeds
the requirements of its inputs and outputs. In the final step, the interfaces are converted to match.
Figure \ref{fig:org5fb76c4} shows all the hardware interface conversions
that HWImg may insert to match interfaces. Fan-In and Fan-Out conversions take tuples that have 
been implemented as multiple streams, and synchronize them to one stream (and fan-out does the opposite). 
Serialize and
de-serialize conversions perform conversions between rates and vector widths (to convert between valid points,
as shown in figure \ref{fig:orgf378fa4}). Finally, Static interfaces are promoted to Stream in Stream pipelines.

One unique feature of our mapping approach is that conversions are only inserted if needed - HWTool does
not force
each intermediate to be converted to some canonical interface. 

Mapping functions always have the option 
to return a Rigel2 module with the same interface as its input, avoiding any conversions. We think this is one reason our
relatively simple mapping approach works well in practice.

\section{Implementation}
\label{sec:org1b00d1f}

To evaluate the efficiency of our pipelines, we synthesized the Verilog generated by HWTool for the Xilinx
Zynq UltraScale+ ZU9-EG, a mid-range FPGA with attached ARM processor and AXI memory system.
To synthesize our Verilog code into an FPGA design, we used Xilinx's Vivado 2018.2, and recorded the area 
requirements and
clock frequencies reported by this tool. To check the correctness of our pipelines, we simulated each
pipeline using Verilator 4.034, a leading open source Verilog simulator, and verified that each pipeline
produced exactly the same output as a verified reference image. Our Verilator test-bench includes simulation of
the AXI memory interfaces and memory system on the ZU9. Cycle counts were recorded from Verilator 
simulation runs. 

\section{Evaluation}
\label{sec:org49d8143}

To test the correctness, scope, and quality of designs produced by HWTool, we implemented a number
of full-scale image processing pipelines in HWImg, and used HWTool to map them to hardware. We then
synthesized this hardware for a Xilinx UltraScale+ FPGA. We tested the following pipelines:

\convolution~performs an 8x8 convolution on a 1080p image. This is our simplest pipeline, 
but it is a challenging test
of hardware quality: it does relatively little compute compared to the other tests, so any unnecessary
hardware overhead produced by the compiler will be apparent.

\stereo~compares 8x8 pixel overlapping patches between two images, and returns the patch match 
with the lowest
Sum of Absolute Difference (SAD) cost. This pipeline could be used to compute depth from
stereo, or to perform block matching for compression. For this test, we perform 64 block matches 
on a 720x400 pixel image.

\flow~computes dense Lucas-Kanade optical flow on a pair of images \cite{lk}. Unlike \stereo,
Lucas-Kanade finds matches between patches using a least-squares solver, which involves computing image
gradients and solving a small linear system. This pipeline tests how this common class of algorithm in computer
vision performs in our system.

\descriptor~computes a simplified sparse Histogram of Gradients (HoG) style feature descriptor.
This pipeline tests two key features of HWTool. 
Descriptors are only computed at Harris 
corner points, so this pipeline performs computations on sparse,
bursty data-dependent streams. 
Second, this pipeline uses floating-point math to compute and scale the 
high-dynamic-range histograms. Verilog does not support float natively, so we used Berkeley's HardFloat 
library \cite{hardfloat}. This demonstrates how HWTool can import external
Verilog code, including complex modules like a floating point divider that has data-dependent latency.

\subsection{Scheduling Range \& Efficiency}
\label{sec:org395a954}

First, we will evaluate the range of schedules supported by HWTool,
and their resulting efficiency.
To understand efficiency, this section will evaluate the hardware resources needed for each schedule.
Key metrics will the number of FPGA Configurable Logic Blocks (CLBs), Block RAMs (BRAMs), and 
Digital Signal Processing 
(DSP) blocks \cite{xilinxarch}. 
For all our results, we disabled usage of DSPs, with the exception of floating point units in \descriptor. Mapping
into DSPs is unreliable, and makes it difficult to compare the relative efficiency of different schedules, because an
inconsistent percentage of each schedule gets mapped to DSPs.

All of our pipelines can attain clock rates between 95MHz-150Mhz on our test FPGA.
We did not spend any time optimizing our designs for clock rate.
From our experience, fixed-function image processing hardware can run at high clocks with additional pipelining,
and this optimization could be applied to our hardware if higher clocks were desired.
For the results in this section, we manually allocated FIFOs. Automatic FIFO allocation will be
evaluated separately in section \ref{sec:org8538dd2}.

\begin{figure}[htbp]
\centering
\includegraphics[width=240pt]{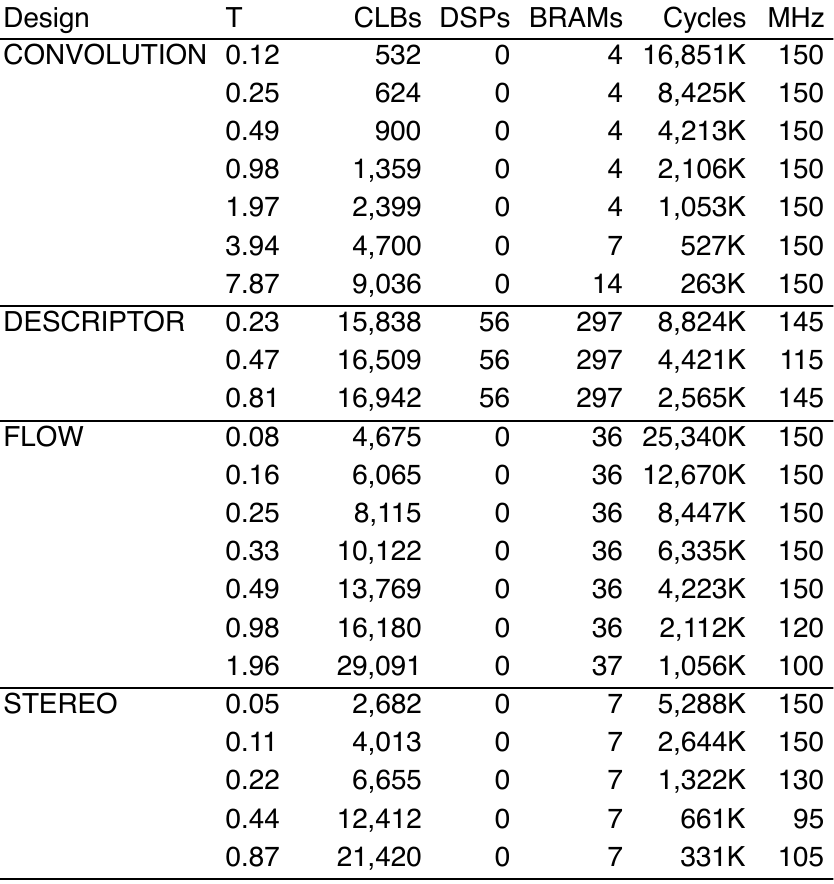}
\caption{\label{fig:org64b1f4f}
We mapped each of our test pipelines to FPGA over a range of throughputs (\(T\)) in pixel/cycle. HWTool swept the full range of schedules for our FPGA, limited on the high end by FPGA fabric size, and on the low end by the smallest array operation.}
\end{figure}

A key features of HWTool is that it can support a wide range of schedules for each pipeline with no
manual annotations required.
Each schedule we consider will be specified by its throughput (\(T\)) in pixels per cycle. For example,
to process a 1080p image, \(T=1\) would requires \(2,073,600\) cycles, and \(T=2\) would requires \(1,036,800\). Ideally,
the amount of hardware required should scale with \(T\): doubling the throughput should double the hardware
resources required.

\subsubsection{Schedule Range}
\label{sec:orgddcf1b5}
To test the range of schedules supported by HWTool, we took each of our pipelines, swept
a range of throughputs, and recorded the resources required, which are given in table \ref{fig:org64b1f4f}. 
The range of valid \(T\)'s is bounded on the high end by the amount of FPGA compute
and bandwidth available, and on the low end by the minimum size of arrays (HWTool does not
share logic between operations). 

HWTool successfully mapped to the full valid throughput range
for this FPGA. To collect these results,
we had HWTool generate designs with \(T\)'s at powers of two (i.e., 0.25, 0.5, 1, 2,\ldots{}). HWTool does not produce hardware at exactly
the \(T\) requested, however this is not a failure: all vector operations must be rounded
up to next highest factor of the array size, which may result in faster hardware than requested.

\subsubsection{Schedule Efficiency}
\label{sec:org433137e}
\begin{figure}[htbp]
\centering
\includegraphics[width=240pt]{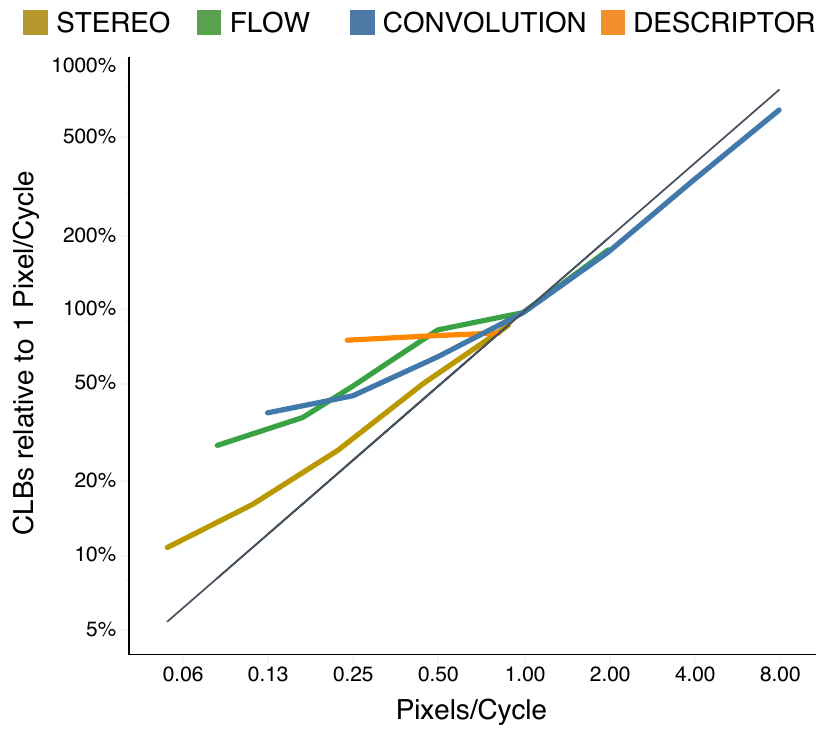}
\caption{\label{fig:orga79a389}
To measure schedule efficiency, we show how our designs scale with throughput, by normalizing CLB resources relative to T=1. most pipeline scale nearly linearly (shown by the black line). Compute-heavy pipelines like \stereo~ and \flow~ scale the best, and the low-compute, sparse \descriptor~ does not scale at all.}
\end{figure}

Directly comparing each schedule to a manual design would be difficult, due to the time to implement
each design in RTL.
Instead, as a measurement of efficiency, we will look at the \textit{scaling} of hardware required for different 
schedules.
Ideally, we should see a linear relationship between \(T\) and hardware resources.
If we see a linear relationship, it is still possible that all our designs have some fixed percentage overhead relative to manual designs,
but this would suggest that our designs are efficient.

To assess scaling, we plotted the \textit{relative} hardware resources required for each schedule
in figure \ref{fig:orga79a389}. 
This plot normalizes
the hardware resources for each schedule to be relative to the resources for \(T=1\).
\(T=1\) is usually the simplest schedule (all array operations to compute one pixel are unrolled), so it 
serves as a
good baseline.

We see that resource scaling is nearly linear
for most pipelines. In general, we expect pipelines that perform more compute relative to buffering
 and control
to scale nearer to linear, and this is seen in the results. 
\stereo~is our simplest compute-heavy pipeline, and it scales nearly
linearly, with \flow~ and \convolution~ doing slightly worse.
 \descriptor~performs computations on sparse feature points, 
so the amount of actual
compute it requires is very small compared to the other pipelines. Because if this, it barely scales at all.

\subsection{Comparison to Manual Scheduling}
\label{sec:orgc09517d}

\begin{figure*}
\centering
\includegraphics[width=516pt]{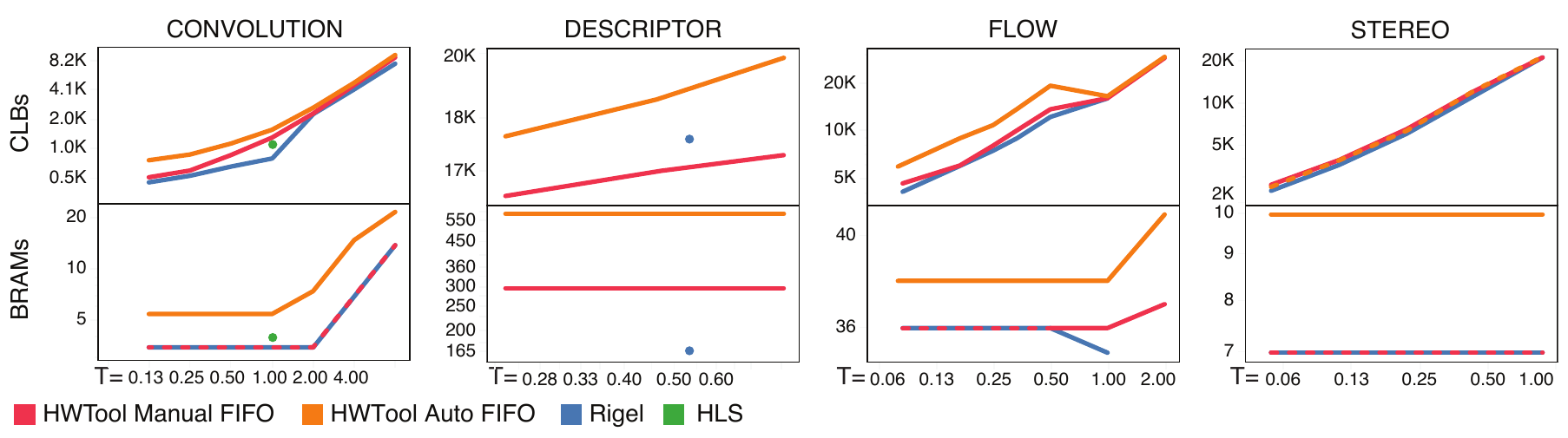}
\caption{\label{fig:orga37db55}
HWTool's auto-scheduled designs compared to Rigel's manual schedules. HWTool's auto-schedular with manual FIFO allocation performs similarly to Rigel, with the exception of \descriptor, where a larger FIFO was intentionally chosen to allow that pipeline to be mapped to a range of throughputs. HWTool's automatic FIFO allocation has BRAM overhead due to excessively conservative handling of bursts. \HLSSTUFF{We also compared HWTool to a HLS compiler on \convolution, and see that its performance is similar.}}
\end{figure*}

Next, we compared HWTool's auto-scheduled designs to manually-scheduled FPGA designs from Rigel
 \cite{rigel}. Since both HWTool and Rigel use the same library of hardware generators,
they have similarities in the hardware they generate.
However, compared to HWTool, the Rigel designs are based on careful manual sizing 
and hardware unit choice. So, any inefficiencies
in hardware sizing or unnecessary conversions introduced by HWTool should be apparent.
For this section, we manually allocated FIFOs similarly to the Rigel designs, to eliminate this as
a factor, and synthesized the Rigel designs at the same clock rates as the HWTool designs.

We present HWTool CLB and BRAM counts relative to Rigel in figure
\ref{fig:orga37db55}. In general, the results track closely, with \convolution, \flow, and \stereo~ being almost identical,
 only differing a small amount in control logic. \descriptor~shows a larger difference between HWTool and Rigel:
for this pipeline, we manually enlarged the sparse Filter operator's FIFO so that the pipeline could perform well
across a range of throughputs (which Rigel did not support). As explained in
section \ref{sec:org3e5493f}, data-dependent operations like the sparse Filter must be annotated manually
based on their performance on real datasets.

\subsection{Automatic FIFO Allocation}
\label{sec:org8538dd2}

Next, we compared HWTool's automatic FIFO allocation (sec. \ref{sec:org2d12b12}) with manual allocation.
Using automatic allocation enables our system to compile to hardware
with no annotations, however it has some overhead as
seen in figure \ref{fig:orga37db55}. 

\convolution, \stereo, and \flow~
have small overheads in BRAMs and CLBs with automatic allocation
relative to manual allocation. The overhead mainly comes
from bursty pad and crop operators that zero-pad the image's boundaries. Hiding these bursts is not actually
necessary, because these operators are attached to AXI DMAs, which have sufficient bandwidth to service the bursts.
Our simple FIFO allocation scheme does not exploit this, but the manual designs do. In \descriptor,
two extra delay buffer slots caused the data-dependent Filter FIFOs (set at 2048 by the user) to jump to the next largest
ram size, doubling the BRAM count.

HWTool allows the user to manually override how hardware is generated, so these overheads
could be easily eliminated with a few annotations. However, we decided to include
these results `as is,' because we think they are representative of the overheads users
may encounter if they spend no time optimizing their pipelines.

\subsection{Comparison to High-Level Synthesis}
\label{sec:org9afad1c}

Finally, we compared HWTool to an industry standard High-Level Synthesis (HLS) compiler on the 
\convolution~ pipeline (fig. \ref{fig:orga37db55}). Unlike HWTool, each HLS schedule variant requires schedule annotations and significant code
re-organizations, so we mapped only \(T=1\). To match HWTool,
the HLS pipeline was also synthesized at 150Mhz. The results from HWTool and HLS are similar (1,153 CLBs for HLS compared to HWTool's 1,359),
providing further evidence that HWTool does not introduce excessive overhead in its mapping process.

\section{Prior work}
\label{sec:org04f11ab}

Replacing Verilog has been an active area of research.
C-to-gates High-Level Synthesis (HLS) tools such as Xilinx's Vivado or Mentor Graphic's Catapult
take blocks of C++ code and turn them into functionally-equivalent hardware modules \cite{vivado,catapult}.
HLS tools have been used successfully in industry on a number of products, and share some 
similarities with HWTool in that they take C++ code as input and abstract the details
of low-level hardware design. However, instead of our embedded language approach, HLS tools take
the C++ language itself as input.

Scheduling C++ onto custom hardware is a difficult problem, so HLS tools require the user to provide detailed annotations to
guide how code should be mapped (such as the parallelism of loops, 
and RAM allocation). From our experience, getting good quality out of HLS requires significant
code rewrites and knowledge of both hardware design and the HLS compiler.
While HLS tools have definitely increased the productivity of designers, we think the limitations of current
 HLS tools make them more like
`Verilog in C' instead of a tool that allows non-experts to map C++ to hardware.

The novel hardware design languages Magma (embedded in Python) and Chisel (embedded in Scala)
\cite{magma, chisel} have started to gain industry adoption. 
These languages are intended for direct low-level specification of the hardware, so they instead
serve as a direct replacement for Verilog, not as a high-level language mapping tool like HWTool.

Halide is a full-featured
image processing DSL that has been used for projects in industry \cite{halide}. Prior work has show that a
subset of Halide can be mapped to hardware using HLS compilers as a backend \cite{jingpu}.
We think Halide is a promising front-end for hardware, however we decided not to use
it because it does not currently provide support for analyzing and optimizing sparse workloads, which
were important use cases for us, and it would be difficult to have Halide code closely integrate with
existing hardware blocks in Verilog.

\section{Discussion}
\label{sec:orge8f03d4}

We presented HWTool, a novel framework for mapping high-level C++ code to hardware with
no scheduling annotations. We demonstrated that the scope of HWTool can
map complex pipelines like Lucas-Kanade optical flow, depth from stereo, and feature descriptors.
Our automatically-generated designs are on average only 11\% larger
(for manual FIFO allocation) and 33\% larger (for automatic FIFO allocation) than hand-optimized designs\HLSSTUFF{, and competitive with HLS}.

We are excited that HWTool may bring more structure, consistency,
and ease to the process of mapping algorithms to hardware. An open problem for HWTool, and the hardware design community in general
is the lack of high-quality open source hardware libraries.
We are encouraged by recent progress in this area, particularly around RISC-V, however
in general hardware does not yet have a culture around sharing open source code \cite{rocketchip, bjump}.
We are also excited about new research our framework may enable by breaking the difficult hardware mapping 
problem into smaller composable units. For example, each of our individual operators, mapping functions, and Verilog modules
is simple enough that each interface and schedule variant could be automatically synthesized from a single behavioral
description and formally verified, whereas this is difficult at the scale of full pipelines.

\bibliographystyle{ACM-Reference-Format}
\bibliography{hwtool}
\end{document}